\documentstyle[12pt,aaspp4]{article}
  
\begin{document}
   
\title{Detecting Compton Reflection and a Broad Iron Line 
in MCG$-$5-23-16 with $RXTE$}
    
\author{K. A. Weaver}
\affil{Johns Hopkins University, Department of Physics and Astronomy,
Homewood Campus, 3400 North Charles Street, Baltimore, MD 21218-2695;
e-mail: kweaver@pha.jhu.edu}
\and
\author{J. H. Krolik}
\affil{Johns Hopkins University, Department of Physics and Astronomy,
Homewood Campus, 3400 North Charles Street, Baltimore, MD 21218-2695}
\and
\author{E. A. Pier}
\affil{NASA Goddard Space Flight Center, Laboratory for High Energy 
Astrophysics, Greenbelt, MD 20771}
      
\begin{abstract}

We report the detection with the Rossi X-ray Timing Explorer
of a Compton reflection signature 
in the Seyfert galaxy MCG$-$5-23-16.  $RXTE$ also resolves the Fe 
K$\alpha$ fluorescence line with FWHM $\sim$ 48,000 km s$^{-1}$.  This 
measurement provides the first {\it independent} confirmation 
of $ASCA$ detections in Seyfert galaxies of broad Fe K$\alpha$ 
lines that are thought to be the  
signature of emission from the inner regions of an accretion disk orbiting
a black hole.  Under the assumption that reflection arises from 
an isotropic source located above a neutral accretion disk, and using a
theoretical model that accounts for the dependence of the reflected spectrum  
on inclination angle, we derive a 90\% confidence range 
for the disk inclination of $i$ = 50$^{\circ} - 81^{\circ}$. 
The large inclination is consistent
with that expected from the unified model for MCG$-$5-23-16
based on its Seyfert 1.9 classification.  If we assume that 
the high-energy cutoff in the incident     
spectrum lies at energies larger than a few hundred keV, then 
the equivalent width of the Fe K$\alpha$ line is 
much larger than predicted for the amount of reflection.
This implies either an enhanced iron abundance, 
a covering factor of reflecting material $c_f > 0.5$, 
or a cutoff in the incident spectrum at
energies between $\sim$60 and $\sim$200 keV.
      
\end{abstract}
	  
	   
\section{Introduction}

At the heart of an active galactic nucleus lies the putative
black-hole ``monster'' that supplies its power.  Much 
of this power emerges in the form of X-rays, which are in turn efficient
probes of the immediate regions surrounding the nucleus.  The fuel
supply for the massive central 
engine is thought to arise from a disk of dense, accreting matter  
with column densities $N_{\rm H}$ well in excess of $\sim 10^{24}$ cm$^{-2}$.   
In the presence of such matter, the reflected X-ray spectrum has
two distinct features.  One is the Fe K$\alpha$ fluorescence emission line at
6.4 keV (Fabian et al.\ 1989, George \& Fabian 1991).
The other is Compton reflection, a broad hump centered on a few tens of keV
with a shape governed by photoelectric absorption at low energies and
by Compton recoil and the declining scattering cross-section at high
energies (Lightman \& White 1988; Guilbert \& Rees 1988). 

The ubiquity of Fe K$\alpha$ lines in Seyfert galaxies is well established. 
Recent results from the Advanced Satellite for Cosmology 
and Astrophysics ($ASCA$)  
suggest that many of these lines in Seyfert 1 galaxies exhibit relativistic 
profiles that are caused by photons emerging near a black hole, and thus 
they most likely originate in an accretion disk 
(Nandra et al.\ 1997).  Whether Compton reflection is as common   
is not known.  Although hard X-ray observations   
with $Ginga$ show that Compton reflection is
clearly present in the {\it composite} spectrum of eight bright Seyfert 
1 galaxies (Pounds et al.\ 1990), the level of significance
of this feature and constraints for
theoretical reflection models are poor for most {\it individual} 
objects in the $Ginga$ sample (Smith \& Done 1996, Nandra \& Pounds 1994). 
The Rossi X-ray Timing Explorer ($RXTE$)
provides the large bandpass coverage and collecting 
area necessary to allow the unambiguous detection of
Compton reflection in many individual Seyfert galaxies
for the first time.

MCG$-$5-23-16 is an X-ray bright active galaxy  
with an optical classification of Seyfert 1.9 
(Veron et al.\ 1980).  In the 20 years since
its discovery, 
its 2 to 10 keV flux has varied by a factor of 4 from 
a high state of $\sim 8 \times 10^{-11}$
ergs cm$^{-2}$ s$^{-1}$ in 1978 (Tennant 1983), to 
a low state of $\sim 2 \times 10^{-11}$ ergs 
cm$^{-2}$ s$^{-1}$ in 1989 (Nandra \& Pounds 1994), 
to a high state of $\sim 9 \times 10^{-11}$ ergs cm$^{-2}$  
s$^{-1}$ in 1996 (this work).
In its high-flux state, MCG$-$5-23-16 is among the five 
brightest Seyfert galaxies in hard 
X-rays.  $ASCA$ observations show the Fe K$\alpha$ line 
to be significantly broadened with a FWZI of $\sim$120,000 km s$^{-1}$ 
and an unusually complex profile (Weaver et al.\ 1997, hereafter W97).  
The signature of Compton reflection has not before been 
detected in this galaxy. 

Here we present the first results for MCG$-$5-23-16 from $RXTE$. 
This work showcases the data quality, and more importantly, the  
power of using $RXTE$ for purposes of X-ray spectroscopy 
studies of active galaxies. 
This paper describes early results from a comprehensive spectral 
and variability study of MCG$-$5-23-16 that involves simultaneous $ASCA$ 
observations.  The $ASCA$ results and a joint analysis 
of the $ASCA$ and $RXTE$ data will be presented in a future paper.

\section{Observations and Data Selection}

$RXTE$ was launched on 30 December 1995 and carries three scientific 
instruments: the Proportional Counter Array (PCA),
the High-Energy X-ray Timing Experiment (HEXTE), and the
All-Sky Monitor (ASM).  Here we present data from  
the PCA, which consists of five collimated (1$^{\circ}$ FWHM)
proportional counter units (PCUs) that contain three multi-anode
detector layers with a mixture of xenon and methane gas.  
The PCUs are numbered 0 through 4.  Each has a bandpass of 2 
to 60 keV, a geometric collecting area of $\sim$1,400 cm$^2$,
and a FWHM energy resolution of $\sim$ 8\% at 6.6 keV.
See Glasser, Odell, and Seufert (1994) for a detailed 
description of the PCA.

MCG$-$5-23-16 was observed with $RXTE$ from November 27 to 30 1996,
for about $\sim$100 ks.   There was an overlapping observation 
with $ASCA$ that began about half way through 
the $RXTE$ observation on 29 November 1996 and lasted for
approximately 35 ks.   We do not discuss the most 
recent $ASCA$ data here  
except to use the observed photon index and 
absorbing column density  
(Weaver et al., in prep) to constrain the  
$RXTE$ fits.

To maximize the signal-to-noise ratio between 2 and 10 keV, 
we accumulate photon events from the top xenon/methane layer
of the PCA.  We also use only PCUs 0 through 3; data from
PCU 4 is discarded because this detector is turned on 
for a smaller fraction of the on-source time due to   
breakdown.  Data is discarded for times of Earth occult
(when the Earth elevation angle is less than 10$^{\circ}$), 
passage of the satellite 
through the South Atlantic Anomaly (SAA), and when 
there is severe electron contamination.  This yields a total   
integration time of $\sim$80 ksec.  For the total spectrum, 
the small statistical errors on the data points 
combined with systematic uncertainties in the detector
response matrix cause large values of $\chi^2$.  Also,
the background subtraction is not yet accurate for the complete  
observation due to small variations in the internal 
background that are not accounted for in the current background  
model.  We therefore use an 11 ksec exposure that
coincides with the $ASCA$ observation
on 29 November 1996, and is accumulated
from $9.188 \times 10^7$ s to $9.191 \times 10^7$ s $RXTE$ mission
elapsed time.  For this snapshot, Poisson statistics are an 
acceptable description of the point to point variations in 
the data, the systematic errors
have a minimal impact on our results because 
they are comparable to the statistical 
errors, and the background 
model works appropriately. 

The PCA detector response matrix was generated by adding together
matrices for PCUs 0, 1, 2, and 3 that were supplied by the $RXTE$
Guest Observer Facility (GOF) at the Goddard Space Flight Center.  The 
individual matrices represent the most up-to-date PCA calibration and
were created 26 August 1997 using {\it pcarmf} v2.2.1.  At the time of
this writing the in-flight energy resolution of the PCA detectors has
not been precisely determined, but based on spectral fits
to line emission from Cas-A, the most likely value has been measured to within 
$\sim$1\% (Keith Jahoda, 1997 private communication).  We therefore use
two response matrices that bracket the possible values of the energy 
resolution.  The v2.2.1 matrix has a FWHM resolution  
of 7.5\% at 6.6 keV (resolution factor of 0.75) as the default value.
The two response matrices that bracket the most appropriate value 
for Cas-A have multiplicative factors of 1.07 and 1.13 on the default 
resolution yielding effective resolution
factors of 0.8 and 0.85, or 8.0\% and 8.5\% FWHM resolution at 6.6 keV.
These response matrices are labeled M8 and M85 in the
spectral fits discussed below.  In the work presented here, the
uncertainty in the resolution only affects the significance
of detection of the broad Fe K$\alpha$ line.
Power-law fits to the spectrum of the Crab with the v2.2.1 
response matrix yield residuals less than 2\% at all energies
included in our fits (Keith Jahoda, 1997 private communication).

The amount of background internal to the PCUs varies 
according to the orbital environment of the satellite.
There are no detectors offset from the science axis to 
allow a simultaneous measurement of the background,   
and so the standard procedure that has been developed 
by the $RXTE$ GOF is to model the internal and cosmic 
background.  The background for MCG$-$5-23-16 was modeled in 
February 1997 with the $pcabackest$ program.  The 
internal background was predicted from the   
particle and activation models that have been derived  
from Earth-occult data.
The particle model is based on the value of the 
housekeeping parameter Q6, which tracks  
the instantaneous particle flux by measuring the rate of events that
trigger exactly 6 of the lower level discriminators in each PCU.
The activation model estimates the additional  
background that is present in the detectors
after the satellite passes through an SAA. 

After background 
subtraction, the PCA count rate for MCG$-$5-23-16 is 
$\sim$40 counts s$^{-1}$.  The 11 ksec spectrum contains
$\sim 4.4 \times 10^5$ photons.  We estimate that
the current background model is reliable
up to energies of 30 keV and ignore pulse-height channels 
above this energy.  Channels below
2.8 keV are ignored because of remaining calibration
uncertainties involving partial charge 
collection (Keith Jahoda, 1997 private communication).
Data are modeled with the $XSPEC$ spectral fitting 
package and $\chi^2$ statistics are used.

\section{Spectroscopic Results}

MCG$-$5-23-16 has a large absorbing column of $N_{\rm H}$ 
$\sim$2 $\times10^{22}$ cm$^{-2}$.  This column is measurable 
with the PCA, but because of ongoing 
calibration issues at the lowest energies, 
a systematic error may exist in this measurement 
that would affect our spectral results.
Therefore, we use the information that the simultaneous $ASCA$ data 
provide near energies of 2 keV to place an  
additional constraint on our spectral modeling.
In all cases, we fix the absorbing column density at 
$N_{\rm H} = 1.65 \times 10^{22}$ cm$^{-2}$ (Weaver et al., in prep),
which was obtained by fitting the most recent 
(simultaneous with $RXTE$) $ASCA$ data from $1-4.5$ keV 
with an absorbed power-law model.  
This column is identical to 
that measured during the 1994 observation 
($N_{\rm H} = 1.62\pm0.2
\times 10^{22}$ cm$^{-2}$; W97).

The results of spectral fitting are listed in Table 1.
We use three models to describe the spectrum: a 
power law 
(N$_{\rm E} \propto E^{-\Gamma}$), a power law plus a Gaussian, 
and a Compton reflection model plus 
a Gaussian.  The reflection model is the $pexrav$ model in $XSPEC$,
which is an exponentially cut-off
power-law spectrum reflected from a disk of neutral material  
(Magdziarz \& Zdziarski 1995).  The free
parameters are the photon index of the primary 
power-law spectrum ($\Gamma$), the cutoff energy of the 
primary power-law spectrum ($E_c$), 
the abundance of heavy elements ($Z$),
the relative amount of reflection compared to the 
directly-viewed primary spectrum ($R$), the inclination 
angle of the disk normal to our line of sight ($i$),  
and the power-law normalization ($A$).
The quantity $R$ is equal to 1 for the case of an
isotropic X-ray source above a disk with 
a covering factor $c_f$ = 0.5 ($\Omega$ = 2$\pi$) as 
viewed from the X-ray source. 

The amount of reflected flux  
depends on $i$ and $Z$, with larger inclinations and/or 
larger abundances producing less 
reflection (George \& Fabian 1991).
In addition, because of the contribution to the reflection
hump by down-scattering of high energy photons,
the amount of reflection depends on the cutoff 
energy of the primary power-law spectrum.
If $E_c$ is larger than a few hundred keV, there 
is no significant reduction in the amount of
reflection at $<$30 keV in comparison to a pure
power-law spectrum; however, if  
$E_c$ is less than a few hundred keV, 
there are fewer primary photons that can scatter to
energies below 30 keV.  To test the effect of 
varying $E_c$, we examine 
models with $E_c$ = 60 keV, 200 keV, and 500 keV.   
Cutoff energies much less than 60 keV are unlikely 
because they provide poor fits unless $\Gamma$ 
$<$ 1.75, which is outside the range of 
indices allowed for the intrinsic spectrum by the simultaneous 
$ASCA$ data (Weaver et al., in prep).  A value of
$E_c$ = 200 keV approximates the cutoff implied
by fits to non-simultaneous OSSE and $Ginga$ data
for MCG$-$5-23-16 (Fig.\ 7 in Zdziarski, Johnson,
\& Magdziarz 1996) and is similar
to the mean cutoff energy for
radio-quiet Seyfert 1 galaxies (Gondek et al.\ 1996).
The upper limit of 500 keV is chosen simply to have 
no appreciable effect in the PCA bandpass.
We also assume Solar abundances and 
allow only one of the parameters $R$ and $i$ 
to be free for a given fit.

For all models, the spectral parameters 
and statistical error bars that are derived using
response matrices M8 and M85 (\S 2) are similar, if not
identical, and so we list only the numbers derived 
using M8 in the table (columns 2 through 8).
The difference between M8 and M85 does matter when assessing 
the significance of detection of the broad Fe K$\alpha$ line.
We therefore list the values of $\chi^2/\nu$, the 
F-statistic, and the probability of exceeding F 
for both M8 and M85 in columns 10, 11, and $13-16$.
For fits that include reflection, we also list three
best-fit values for the parameters that are most 
dependent on $E_c$.  These are the iron line 
equivalent width (W$_{{\rm K}\alpha}$, column 5) 
and $i$ or $R$ (columns 7 and 8). 
The top entry corresponds to $E_c$ = 500 keV, the
middle entry corresponds to $E_c$ = 200 keV, and the 
bottom entry corresponds to $E_c$ = 60 keV.  

Models that do not include reflection 
provide formally unacceptable fits. 
An absorbed power law with $E_c$ placed above the PCA 
bandpass provides an exremly poor 
fit (Table 1, fit 1) with $\chi^2/\nu > 9$ for 57 d.o.f. 
The poorness of this fit is illustrated in Figure 1, which 
shows the PCA data and the power-law model folded through the instrumental
response.  The residuals for the fit are plotted in the bottom panel. 
Adding a narrow Gaussian to the model provides   
a large improvement (Table 1, fit 2, Fig.\ 2a), but the fit is still 
statistically unacceptable with $\chi^2/\nu$ $\sim$ 1.5 for 55 d.o.f.  
The probability of obtaining such a large value of $\chi^2$ by chance
is only $\sim$0.01.  If the line is allowed to be broad 
(Table 1, fit 3, Fig.\ 2b), 
the fit is improved significantly for response M8, but not for 
M85, as indicated by the small F-statistic and large probability of 
exceeding F in columns 14 and 16.   This fit, however,
remains formally unacceptable for both matrices (columns 10 and 11) 
with a probability of $\sim$0.02
for obtaining such large values of $\chi^2$ by chance.  

Models that include reflection provide  
formally acceptable fits (Table 1, fits $4-7$).
To examine how significant this improvement is 
compared to models that include only the iron line, 
we compare the fit with a power law and a broad Gaussian (fit 3),
to a fit with a power law plus reflection and
a narrow Gaussian (fit 4, Fig.\ 2c). 
For the fit with reflection, the decrease in $\chi^2$
is large compared to the fit without reflection; $\Delta\chi^2$ =
14 for M8 and $\Delta\chi^2$ = 21 for M85.
F-statistic values are not listed for fit 4 because the
F-statistic cannot be used in this case to accept or
reject the hypothesis that reflection provides a better fit to
the data than a broad line.  Instead, we calculate
the ratio of likelihoods
(Edwards 1972) for the reflection
plus narrow Gaussian model versus the power law plus
broad Gaussian model.  This ratio is defined as L$_1$/L$_2$ =
exp[${\Delta\chi^2}/2$] for two models.  We find that
reflection plus a narrow Gaussian is 1,100 to 36,000
times more likely to be the correct description of the
data than a power law and broad Gaussian, assuming that the
errors are normally distributed.  

We next examine whether a broad line is detected when 
reflection is included in the model (Table 1, fits 5 to 7).
To test the range of allowed parameter space, we assume
different values for $i$ and $R$.  For fits 5 and 6, we fix  
$i$ at values of $50^{\circ}$ and
$87^{\circ}$, which cover the range of most
likely inclinations for a neutral accretion
disk measured from $ASCA$ studies of
the Fe K$\alpha$ line profile (W97).  Angles 
less than $50^{\circ}$ are the least preferred by 
the $RXTE$ data because they yield poor fits
unless $E_c$ is as low as $\sim30-50$ keV, 
which we have already ruled out since this requires the 
intrinsic spectrum to be too flat. 
For $i = 50^{\circ}$, $R$ is small and ranges from 
0.36 to 0.80 implying covering factors of $c_f = 
0.18 - 0.40$; however, for $i =  87^{\circ}$, 
$R$ is large and ranges from 2.9 to 5.9, implying
unphysical covering factors of $c_f > 1.0$.
If we instead fix $R$ at 1.0 (fit 7),
then $i$ ranges from 63$^{\circ}$ to 81$^{\circ}$
(best-fit values) with a 90\% confidence range 
of 50$^{\circ}$ to 87$^{\circ}$.
Our results clearly show that $R$, $i$,
and $E_c$ are strongly coupled.  For a 
given $R$, smaller inclinations and
W$_{{\rm K}\alpha}$ are allowed for  
a low $E_c$, while for a given $i$,   
a low $E_c$ allows more 
reflection and hence a larger $c_f$. 
The confidence contours for fit 7 for the
width of the Fe K$\alpha$ line vs.\ $i$ for
$E_c$ = 200 keV are shown in Figure 3.
For all of the above cases, a broad line significantly
improves the fit over a narrow line, as
indicated by the F-statistic $>3.8$.
The probability of exceeding F by chance
ranges from 0.05 for the worst case (and the
poorest overall fit) to $<$0.001. 
For fit 7, the $2-10$ keV flux is $\sim9.5 \times 10^{-11}$
ergs cm$^{-2}$ s$^{-1}$.

Finally, we examine whether the gas that reprocesses
X-rays is ionized by replacing the $pexrav$ model, which calculates 
reflection from a neutral disk, with  
the $pexriv$ model, which calculates reflection from an
ionized disk.  The $pexriv$ model adds the 
free parameters of disk temperature and 
ionization parameter, but yields no improvement
to the fit compared to the neutral 
case.  Also, the best-fitting
ionization parameter is zero. This result is 
consistent with the fact that the peak energy of the 
Fe K$\alpha$ line is approximately equal to that 
expected for fluorescence from neutral iron
(6.4 keV in the galaxy rest frame), 
and rules out significant ionization of the disk.

We conclude that Compton reflection $and$ a broad Fe K$\alpha$ 
line from a dense, neutral reflector
are detected in MCG$-$5-23-16 with $RXTE$.  In an 11 ksec exposure,
the line width is well constrained.
There is also little difficulty in statistically separating
the reflection component from 
the broad iron line, a problem that has plagued the analysis of
data from missions such as $ASCA$ and $Ginga$.

\section{Discussion} 

We have shown that both Compton reflection and a broad Fe K$\alpha$ 
line are detected in the $RXTE$ spectrum of MCG$-$5-23-16.  This is
the first clear detection of reflection in this galaxy, and the broad 
Fe K$\alpha$ line confirms the reality of the broad line in the  
$ASCA$ data (W97).  Detecting a broad Fe K$\alpha$ line with $RXTE$
in a Seyfert 1.9 galaxy provides the first independent
confirmation of $ASCA$ detections of broad lines in these objects.
This also puts to rest the idea that the broad lines seen  
by $ASCA$ are spurious features caused by a ``continuum conspiracy'',
where a complex continuum mimics a broad line.  

The significance of detection of the broad line (i.e.,
having $\sigma > 0.24$ keV) depends somewhat on the response
matrix.  However, even for the worst-fitting reflection model,
a broad line is detected at $\ge$95\% confidence.
For the best-fitting reflection models, a broad line is detected at
greater than 99\% confidence.

Examining the parameter space for reflection models, we tested disk 
inclinations that range from $i$ = 50$^{\circ}$ to 87$^{\circ}$,
as suggested by a detailed study of the Fe K$\alpha$ profile (W97).
The parameters of disk inclination ($i$), the amount of 
reflection ($R$),
and the high-energy cutoff ($E_c$) are strongly correlated.
If we allow the amount of reflection to be a free parameter,
then an angle of 50$^{\circ}$ provides
the poorest fit of the models we tested.  For $E_c$ = 500 keV,
$R$ is small but for $E_c$ = 60 keV, $R$ becomes consistent 
with 1, which is the value expected if the reprocessing gas 
is in the form of a disk.   An angle 
of 87$^{\circ}$ provides the best fit, 
but for this case we find the opposite result: 
for $E_c$ = 500 keV, $R$ is consistent with 1, but for 
$E_c$ = 60 keV, $R$ is quite large and implies 
an unphysical covering factor of neutral reflecting material
of $c_f > 1.0$.  Assuming instead a normal disk geometry  
with $c_f = 0.5$ ($R = 1$), the
fit is statistically comparable to the best-fit, and 
we confirm the above result, namely that 
$i$ can range from 50$^{\circ}$ 
to 87$^{\circ}$ for high-energy cutoffs 
in the primary spectrum ranging from 60 to 500 keV. 

The model parameters derived from $RXTE$ are 
similar to those derived from $ASCA$ (W97). 
For a reflection model with $c_f = 0.5$, 
$\Gamma$($ASCA$) ranges from 1.84 to 2.05
(90\% confidence errors) depending on how the Fe K 
line is modeled, while the mean value for $RXTE$ from 
reflection-model fits is $\Gamma$($RXTE$) = 1.82$\pm$0.10.   
The indices are consistent although not necessarily identical, 
but this is not a problem since the photon index  
varies in this galaxy (W97).  Comparing the Fe K-line parameters
is not straightforward because the 
line is clearly non-Gaussian (W97) and the energy resolutions of the 
experiments differ by a factor of four.
However, the line parameters are consistent for 
a single-Gaussian approximation.
In $ASCA$, the line has E$_{\rm peak}$ 
= 6.37$\pm$0.05 keV, W$_{{\rm K}\alpha}$ $\sim$ 260 eV, 
$\sigma$ $\sim$ 0.5 keV, 
and FWZI $\sim$2.4 keV (W97).  In $RXTE$, the line has
E$_{\rm peak}$ = 6.26$^{+0.09}_{-0.11}$ keV, 
W$_{{\rm K}\alpha}$ = 230$^{+25}_{-22}$ eV, 
$\sigma$ = 0.43 $\pm$0.15 keV, and FWZI $\sim$2.5 keV. 

The inclination of the accretion disk is determined by 
different methods for the two experiments.  For $RXTE$,
it is measured from the amount of Compton reflection, and 
for $ASCA$, it is measured from the shape of the 
Fe K$\alpha$ line.  W97 and Turner et al.\ (1997; hereafter T97)
derive different inclinations from $ASCA$ using different techniques.  
Both authors use a model that consists of a theoretical 
line profile from an accretion disk (a `disk line';
Fabian et al.\ 1989, Laor 1991) 
and assume fluorescence from   
a neutral disk with $c_f = 0.5$; however, 
W97 use a two-component model 
that consists of a disk line and a   
narrow Gaussian at 6.4 keV (which represents emission  
from gas further out), while 
T97 fit the entire line with a disk line.
Additionally, W97 assume an outer disk 
radius of $r_o=400r_g$ ($r_g$ = GM/$c^2$) 
and derive the following three parameters: 
the disk emissivity index, $q$ (r$^{-q}$),
the inner disk radius, $r_i$, and the inclination ($i$).  
Using a different approach, T97 assume $r_i=6r_g$, 
$r_o=1000r_g$, and $q=2.5$ and derive only $i$.
Depending on the choice of Schwarzschild 
or Kerr geometries, W97 find  
$q$ = 5 to 7, $r_i$ = 13 to 22$r_g$, 
$i$ = 52$^{\circ}$ to 89$^{\circ}$, 
W$_{{\rm K}\alpha}$(Gaussian) = 54 to 69 eV,
and W$_{{\rm K}\alpha}$(disk line)  = 198 to 223 eV. 
 Using a Schwarzschild
geometry, T97 find $i = 33^{+11}_{-4}$ $^{\circ}$ and 
W$_{{\rm K}\alpha}$(total) = 362$^{+94}_{-43}$.  It is not possible to 
prove which assumptions are correct because the geometry
of the disk, the emissivity profile 
as a function of radius,    
and even whether or not all of the 
fluorescence arises from an accretion disk,
are not known {\it a priori}.
We conclude that the inclination derived from the
line profile does not necessarily represent 
of the true inclination of the accretion disk.  
The range of inclinations derived from $RXTE$
from the amount of Compton reflection
is 50$^{\circ} - 87^{\circ}$ (90\% confidence),  
in agreement with W97.

One problem for all of these analyses, including 
the current one, is the large equivalent width of the disk line.
For $E_c$ = 500 keV, we find W$_{{\rm K}\alpha}$ $\sim$ 230 eV
with $RXTE$, which is much larger than the   
maximum equivalent width of 160 eV predicted from a neutral  
disk (George \& Fabian 1991).  The W$_{{\rm K}\alpha}$ of the 
disk line can be reduced if
we follow W97 and assume that $\sim$25\% of the iron line 
flux originates from somewhere other than the disk. 
In this case, we find W$_{{\rm K}\alpha}$(disk line) $\sim$170 eV.
This is still a problem, however, because the 
equivalent width for
an edge-on disk is expected to be much smaller than 160 eV.  
In fact, W$_{{\rm K}\alpha}$ predicted for a neutral disk viewed at 
our best-fitting inclination of 81$^{\circ}$ for an incident 
spectrum with $\Gamma$ = 1.9 
is only $\sim$50 eV (George \& Fabian 1991).
Clearly, there is too much flux in the iron line
to be produced by such a disk. 
Large equivalent widths can be produced if the accretion disk is 
ionized, but the $RXTE$ data rule out this possibility.  

Besides the dependence of W$_{{\rm K}\alpha}$ 
on disk inclination and $\Gamma$,  
W$_{{\rm K}\alpha}$ is proportional to the  
iron abundance and the covering factor of reflecting  
material.  If we assume almost complete coverage of
the X-ray source ($c_f$ = 0.99), an equivalent width
as large as 170 eV restricts the disk to be
inclined by no more than 70$^{\circ}$
(George \& Fabian 1991).  For the case of   
$E_c$ = 500 keV,
this inclination is inconsistent with our statistical
lower limit of 73$^{\circ}$ (Table 1).  Thus 
it is impossible to have a self-consistent 
interpretation of the data for $E_c$ = 500 keV
without having a factor of $\sim$3 overabundance of iron. 

For smaller cutoff energies, smaller 
inclinations are allowed.  For $E_c$ = 60 
and 200 keV, the 90\% confidence range of 
inclinations is 50$^{\circ}$ to 81$^{\circ}$, 
which allows equivalent widths 
as large as $\sim$130 eV for a 
neutral disk with $c_f$ = 0.5.  To produce a 
value as large as W$_{{\rm K}\alpha} = 170$ eV, it is 
thus possible to have $c_f$ as small as 
0.65 and still have solar abundances.  If iron 
is overabundant, $c_f$ can be less.
This technique of comparing the relative
fluxes in the Fe K line and reflection hump 
thus allows us to infer the presence of a high-energy cutoff
in the intrinsic spectrum 
and to place approximate limits on $E_c$
of 60 to 200 keV.  Our analysis clearly
demonstrates the model-dependence of disk inclination angle
and covering factor on the shape of the
incident spectrum and provides a 
self-consistent explanation of the data without
invoking unusually large abundances.  

The unified model describes Seyfert 1 galaxies as 
having their nuclear regions viewed preferentially 
at small inclination angles so that the continuum 
source and broad line region are not blocked.
Conversely, this model describes 
Seyfert 2s and intermediate types (i.e., Seyfert 
1.8s and 1.9s)
as having their inner regions viewed preferentially 
edge-on so that the continuum source and broad line region
are blocked by thick clouds.  Within 
the context of this model, the mean inclinations for 
the two types are predicted to be $\sim$30$^{\circ}$
and $\sim$60$^{\circ}$ for 1s and 2s, respectively.  
Seyfert 1 galaxies appear to follow the expected 
trend, with a mean inclination angle of 30$^{\circ}$
(Nandra et al.\ 1997).  On the other hand, Seyfert 2 
galaxies may not follow the expected trend.
T97 derive small angles of 13 to 33$^{\circ}$ for 
four Seyfert 1.9 and 2 galaxies, including MCG$-$5-23-16, 
and conclude that these objects have their inner
regions preferentially seen face on, similar to 
Seyfert 1s.  In contrast, the $RXTE$ data   
imply that the disk in MCG$-$5-23-16 is viewed edge-on. 
The 90\% confidence range of inclinations 
inferred from fits with Compton reflection models, 
50$^{\circ}$ to 81$^{\circ}$, is entirely
consistent with the geometry of the nuclear region 
that is predicted by the unified 
model based on its Seyfert 1.9 classification.
 
Clearly, $RXTE$ provides valuable information
that can be used to constrain the geometry and 
distribution of gas   
near the centers of active galaxies.  Comparing 
results in the literature, we   
find that using $ASCA$ data alone  
to derive disk inclinations from iron line profiles
is ambiguous because there are many unknown 
parameters and the results 
depend crucially on the assumptions that go into 
the modeling.  Although an approximately face-on disk 
can be derived for certain techniques of modeling 
the $ASCA$ data, this interpretation is not 
supported by the $RXTE$ data, which instead favor 
large inclinations.

\section{Conclusion}

Using $RXTE$, we have detected the 
signature of Compton reflection in the Seyfert 1.9  
galaxy MCG$-$5-23-16.  We also confirm the broad 
Fe K$\alpha$ line seen with $ASCA$ with 
FWHM $\sim$ 48,000 km s$^{-1}$.  This measurement 
provides an independent confirmation of the broad line 
that is thought to be the
signature of emission from the inner regions of an 
accretion disk orbiting a black hole.  From spectral modeling,
we derive a 90\% confidence range
for the disk inclination of 50$^{\circ} - 81^{\circ}$,
which is consistent
with the inclination predicted from the unified model for MCG$-$5-23-16
based on its Seyfert 1.9 classification.
The equivalent width of the Fe K$\alpha$ line from the disk 
is at least $\sim$170 eV.  This is 
much larger than that predicted for a highly inclined,
neutral accretion disk unless we infer supersolar
abundances, an unusually large covering factor
of reflecting material, 
or the presence of a high-energy cutoff in
the incident spectrum between 
$\sim$60 and $\sim$200 keV. 
 
\acknowledgements

We wish to thank Keith Jahoda for invaluable help with implementing 
and understanding the PCA calibration.  We also thank the members 
of the $RXTE$ Guest Observer Facility for their help with the 
data extraction.  This research was supported by a NASA Long Term 
Space Astrophysics Grant.

\newpage

\begin{figure}
\plotfiddle{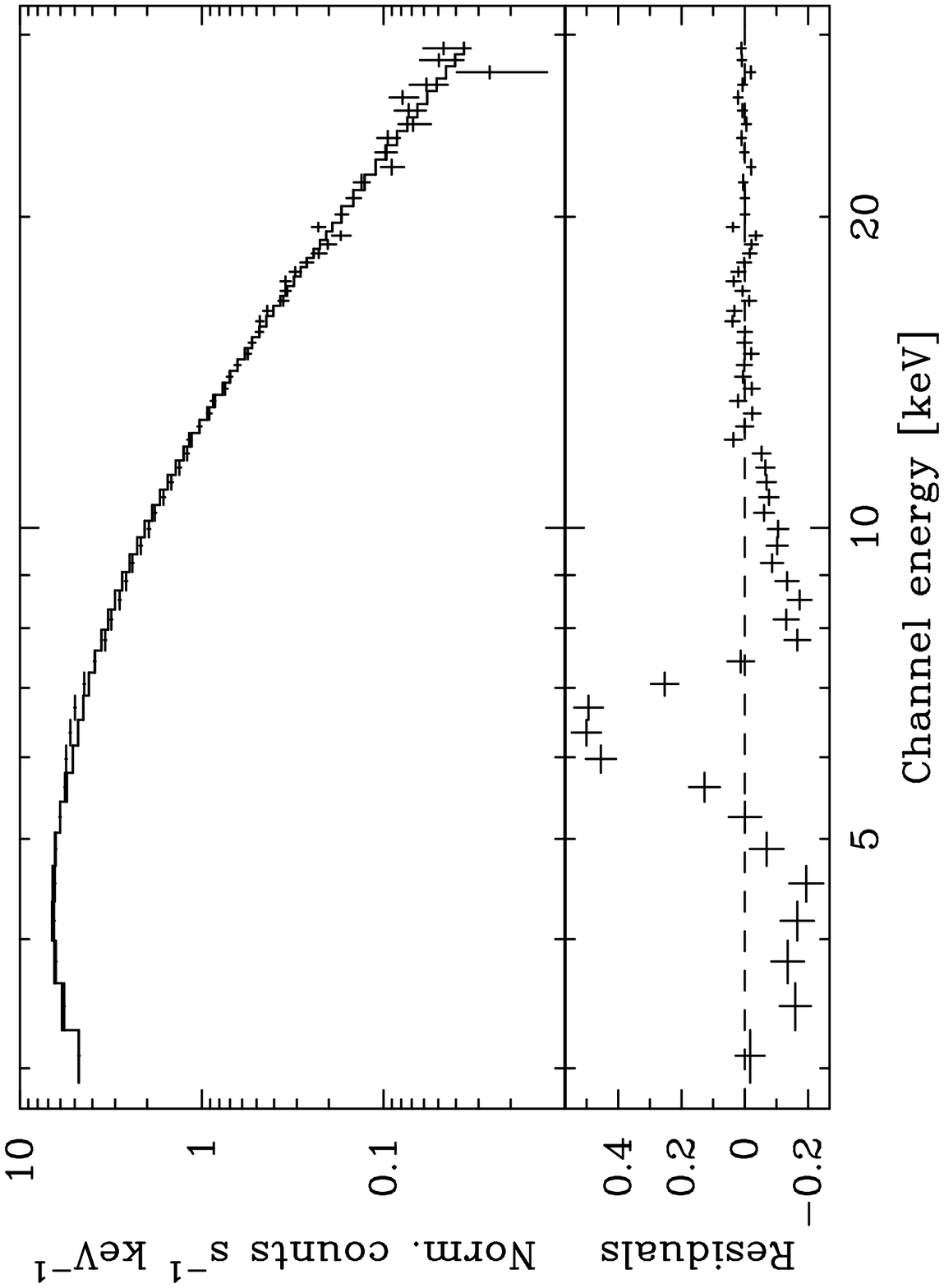}{150pt}{-90}{45}{45}{-185}{255}
\figcaption[ ]{Data from the $RXTE$ PCA and a model that
consists of an absorbed power law (fit 1, Table 1) folded
through the instrumental response.  The
residuals are plotted in the bottom panel.
}
\end{figure}
\begin{figure}
\plotfiddle{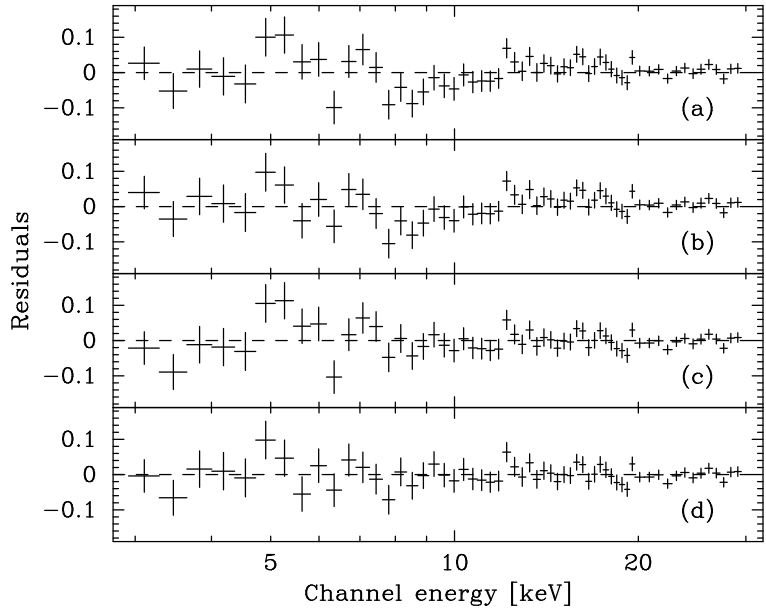}{150pt}{-90}{45}{45}{-185}{260}
\figcaption[ ]{Residuals for fits listed in Table 1 for models  
that consist of (a) an absorbed power law and a narrow 
Fe K$\alpha$ line (fit 2), (b) an absorbed power law and a broad 
Fe K$\alpha$ line (fit 3), (c) an absorbed power law, Compton reflection,
and a narrow Fe K$\alpha$ line (fit 4), and (d) an absorbed  
power law, Compton reflection, and a broad Fe K$\alpha$ line (fit 7). 
All of these fits are for the case with $E_c$ = 500 keV
and response matrix M8.
}
\end{figure}
\begin{figure}
\plotfiddle{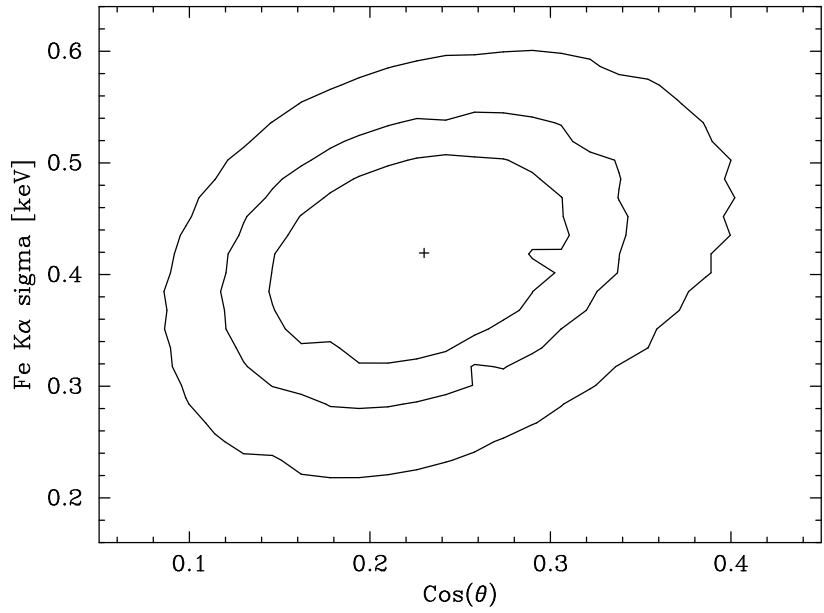}{150pt}{-90}{45}{45}{-180}{275}
\figcaption[ ]{Confidence contours for the Fe K$\alpha$ line width vs.\ 
cosine of the disk inclination angle derived from the 
Compton reflection hump for $c_f$ = 0.5 ($R$=1.0) 
and a primary power-law 
spectrum with E$_{\rm c}$ = 200 keV.
(fit 7, Table 1).
}
\end{figure}

\end{document}